\documentstyle[aps,preprint]{revtex}
\begin{document}

\renewcommand{\baselinestretch}{1.1}
\setlength{\baselineskip}{\baselinestretch\baselineskip}
\newcommand{\edc}{\end{document}}
\newcommand{\disc}{\displaystyle}
\newcommand{\bea}{\begin{eqnarray}}
\newcommand{\be}{\begin{equation}}
\newcommand{\eea}{\end{eqnarray}}
\newcommand{\ba}{\begin{array}}
\newcommand{\banonum}{\begin{eqnarray*}}
\newcommand{\ea}{\end{array}}
\newcommand{\Tr}{\mbox{Tr}}
\newcommand{\beq}{\begin{equation}}
\newcommand{\eeq}{\end{equation}}
\newcommand{\ee}{\end{equation}}
\newcommand{\dsf}{\displaystyle\frac}
\newcommand{\ds}{\displaystyle}
\newcommand{\re}[1]{(\ref{#1})}
\newcommand{\fpisml}{\disc f_{\pi}}
\newcommand{\bra}[1]{\langle{#1}\vert}
\newcommand{\ket}[1]{\vert{#1}\rangle}
\newcommand{\matel}[3]{ \bra{#1}{#2}\ket{#3}}
\newcommand{\vrho}{\vec\rho}
\newcommand{\cald}{{\cal{D}}}
\newcommand{\gpinn}{ G_{ {\pi}NN}}
\newcommand{\gsnn}{ G_{ {\sigma}NN}}
\newcommand{\gomegann}{ G_{ {\omega}NN}}
\newcommand{\grhonn}{ G_{ {\rho}NN}}
\newcommand{\frhonn}{ F_{ {\rho}NN}}
\newcommand{\fomegann}{ F_{ {\omega}NN}}
\newcommand{\ci}{\cite}
\newcommand{\mpi}{{\disc{m}_{\pi}}}
\newcommand{\delb}{ \Delta^B  }

\title{\Large\bf Meson - nucleon vertex form  factors at finite temperature.}

\author{A. Rakhimov$^a$, U. Yakhshiev$^b$ and F.C. Khanna$^c$\\
$^a$Institute of Nuclear Physics, Academy of Sciences,\\
    Uzbekistan (CIS) \\
$^b$Theoretical Physics Labaratory, Institute of Applied Physics,\\
 Tashkent State University
Tashkent-174, Uzbekistan (CIS)  \\
$^c$ Physics Department, University of Alberta,\\
 Edmonton, Alberta T6G 2J1,\\
 and \\
 TRIUMF, 4004 Wesbrook Mall, Vancouver, BC V6T 2A3}

\maketitle

\begin{abstract}
In this paper the dependence of meson-nucleon-nucleon vertex form
factors is studied as a function of termperature. The results are
obtained starting from a zero temperature Bonn potential. The temperature
dependence of the vertex form factors and radii is studied in
the thermofield dynamics, a real-time operator formalism of finite
temperature field theory. It is anticipated that these results will
have an impact on the study of relativistic heavy-ion
collisions as the critical temperature for the phase transition from
hadronic to quark-gluon system is approached.
\end{abstract}

\newpage

\section{Introduction}

 In-medium properties of hadrons and their
interactions is a field of high current interest.
For studying phase transitions and thermodynamical properties of the
nuclear system under extreme conditions, it is essential to determine
the temperature ($T$) and density ($\rho$) dependence of the nuclear force.
Below a critical temperature $T_c$, meson exchange picture
of nuclear physics provides the natural description
for the nucleon -  nucleon ($NN$) interaction.
Quarks and gluons are certainly present in all $NN$ interactions,
but it is not always necessary to take them explicitly
into account especially below a critical temperature
 $T_c$ , when the transition into a quark gluon phase takes
place. At distances smaller than a fermi,  the inner structure of
hadrons is probed since it involves the short distance
or the high momentum components of the wave function. This
 structure is typically taken into
account by including  vertex form  factors.
In hot and dense matter, the  structure of hadrons undergoes changes
which should lead to a modification of the meson - nucleon coupling
constants  or, in general, the form factors.
 The temperature  [1]
and density dependence [2]
of some coupling constants have been investigated recently,
 while that of form factors is still fairly unknown and will be
studied in the present paper for the first time.
For this purpose we shall use the formalism of Thermo Field Dynamics (TFD)
in order to obtain the required temperature and density dependence.

The Thermo Field Dynamics (TFD)  which  was  first suggested by
Takahashi and Umezawa [3]  is  a
 real  time  operator formalism  of  finite  temperature  quantum field
 theory.  In this framework,  all the operator  formalism
 of  quantum  field theory at zero temperature can be extended
 directly to finite temperature and density. Therefore in TFD
 it is possible to  continue  using  Wick's theorem and the
 Feynman diagrammatic approach as in the case of zero temperature
 theory.  An  auxiliary  field  is introduced which leads to  a  field
doublet;  the  propagators  become $ 2*2$
 matrices,  the Feynman  rules are now algebraic operational rules
 in the space of $2*2$  matrices.  Recently,  some  attempts  to
 study  in  medium properties of hadrons at finite temperature
 have been obtained using TFD  [4, 5].

A study of relativistic heavy-ion collisions [6] is expected to lead
us to a proper equation of state of hot and dense nuclear matter. The
behaviour of nuclear
matter as a function of temperature and density is relevant for
investigation in  nuclear physics, astrophysics, cosmology and particle
physics. The phase transition from hadronic system to a quark-gluon
plasma phase and the subsequent hadronisation of the quark-gluon plasma
phase can provide information on the asymptotic freedom as well as
confinement behaviour in Quantum Chromodynamics (QCD). Therefore it is
important to study theoretically the behaviour of the hadronic system
at high temperatures. The present report attempts to study such
a behaviour. In sec. II, a brief review of the
TFD is given. In sec. III,
we calculate the renormalization of meson nucleon vertices
using the Feynman propagators from TFD. The results of calculations
and their discussions will be presented in sec. IV.

\section{Finite temperature formalism for meson and nucleon propagators.}

The Thermo Field Dynamics is a real time operator formalism
of quantum field theory at finite temperature. The main feature of
the TFD is that thermal average of operator $\hat{A}$ is defined
as the expectation value with respect to a temperature dependent
vacuum, $\ket{0(\beta)}$, which is obtained from the regular vacuum by
a  Bogoliubov transformation. Therefore, we have
\be
<\hat{A}>\equiv \Tr (\hat{A}e^{-\beta(H-\mu)})/\Tr(e^{-\beta(H-\mu)})
=\matel{0(\beta)}{\hat A}{0(\beta)}
\ee

where $\beta \equiv 1/k_{B}T$ with $k_{B}$ being the Boltzmann constant, H
is the total Hamiltonian of the system,
and $\mu$ is the chemical potential. Due to the doubling of
all degrees of freedom, every field is represented in  the
thermal doublet form in TFD. For example, the thermal doublet for nucleon
field is
\begin{equation}
\psi^{(a)} (x) \equiv
\left\{
\begin{array}{c}
\psi (x)\\
i \  ^t\tilde{\psi}^\dagger
\end{array}
\right\}
\end{equation}

\noindent where $\psi(x) $ is the ordinary nucleon field and $\tilde{\psi}$
is  the doublet partner of $\psi(x)$. As a consequence
the thermal propagator of a fermion field is a
$2*2$ matrix  defined as
\begin{equation}
i S_{F}^{(a,b)} (x_{1}, x_{2}) = <0(\beta) | T [\psi^{(a)} (x_{1})
\bar{\psi}^{(b)}
(x_{2})] | 0 (\beta) >
\end{equation}

The free thermal propagator  $S_{F}$
  to be used in perturbation theory
can be separated into two parts,
i.e. the usual Feynman part $S_{0}$, propagator at zero
temperature and the temperature dependent
part $S_T$ such that $S_{F}^{(ab)}=S_{0}^{(ab)}+S_{T}^{(ab)}$.
For a fermion with mass $M$, the propagators have the form
\begin{eqnarray}
S_{0}^{(ab)}=(\not p+M)_{(ab)}\left(
\begin{array}{cc}
G_{0}(p^2) & 0\\
0& G_{0}^{*}(p^2)
\end{array}\right)\\
S_{T}^{(ab)}=2\pi i\delta(p^2-M^2)(\not p+M)_{\alpha\beta}\left(
\begin{array}{cc}
\sin^2\theta_{p_0} & \frac{1}{2}\sin 2\theta_{p_0} \\
 \frac{1}{2}\sin 2\theta_{p_0} &-\sin^2\theta_{p_0}
\end{array}\right)
\end{eqnarray}
with
\beq
\begin{array}{l}
G_{0}(p^2)=\frac{1}{p^2-M^2+i\epsilon} \\
\cos\theta_{p_0}=\dsf{\theta{(p_0)}}{(1+e^{-x})^{1/2}}+
\dsf{\theta({-p_0})}{(1+e^{x})^{1/2}},\\
\quad \\
\sin\theta({p_0})=\dsf{e^{-x/2}\theta({p_0})}{(1+e^{-x})^{1/2}}-
\dsf{e^{x/2}\theta{(-p_0)}}{(1+e^{x})^{1/2}},
\end{array}
\eeq
where $a$ and $b$ are Dirac indices and $x=\beta(p_0-\mu)$.
The propagator of the scalar particle with mass $m_B$ is given by
\beq
\delb(p)=\delb_0(p)+\delb_T(p),
\eeq
\begin{eqnarray}
\delb_{0}=\left(
\begin{array}{cc}
D_0(p^2) & 0\\
0&-D_{0}^{*}(p^2)
\end{array}\right)\\
D_{T}=-2\pi i\delta(p^2-m_B^2)\left(
\begin{array}{cc}
\mbox{sinh}^2\phi_{p_0} & \frac{1}{2}\mbox{sinh}2\phi_{p_0} \\
 \frac{1}{2}\mbox{sinh}2\phi_{p_0} &\mbox{sinh}^2\phi_{p_0}
\end{array}\right)
\label{dprop}
\end{eqnarray}
with
\beq
\begin{array}{l}
D_0(p^2)=\frac{1}{p^2-m_B^2+i\epsilon}\\
\mbox{cosh}\phi_{p_0}=\dsf{1}{(1-e^{-|y|})^{1/2}},\\
\quad\\
\mbox{sinh}\phi_{p_0}=\dsf{e^{-|y|/2}}{(1-e^{-|y|})^{1/2}},
\end{array}
\eeq
where $y=\beta p_0$.
 Note that
the propagator of vector mesons would be similar to that of the scalar
mesons except for an additional
factor of $(-g_{\mu\nu}+k_{\mu}k_{\nu}/m^2)$.

\section{Vertex form factors}

 The OBE model of the N-N interaction  [7] includes exchange of several
 mesons. Here we consider the meson-nucleon interactions of $\pi .,
 \sigma ., \omega$- and $\rho$-mesons:
\bea
\ba{l}
{\cal L}_{\pi NN}=-g_{\pi NN}\bar\psi i\gamma_5\vec\tau\psi\vec\varphi_{\pi}\,;
\qquad
{\cal L}_{\sigma NN}=g_{\sigma NN}\bar\psi\psi \varphi_{\sigma}\,;\\
\quad \\
{\cal L}_{\omega NN}=-g_{\omega NN}\bar\psi \gamma_\mu\psi\omega^\mu-
\ds\frac{f_{\omega NN}}{4M}\bar\psi\sigma_{\mu\nu}\psi
[\partial^{\mu}\omega_\nu-\partial^{\nu}\omega_\mu]\,;\\
\quad \\
{\cal L}_{\rho NN}= \left[
-g_{\rho NN}\bar\psi \gamma_\mu\vec\tau\psi\vec\rho^\mu-
\ds\frac{f_{\rho NN}}{4M}\bar\psi\sigma_{\mu\nu}\psi
[\partial^{\mu}\vec\rho_\nu-\partial^{\nu}\vec\rho_\mu\right]\,.
\label{lint}
\ea
\eea

Here $g_{BNN}$ are the meson-nucleon coupling constants and
$f_{\omega NN}$ and $f_{\rho NN}$ are the tensor coupling constants
for $\omega$- and $\rho$- mesons respectively.

The corresponding meson nucleon form factors are usually
defined as:
\beq
\ba{l}
\matel{N(p')}{\Gamma_{\pi}^\alpha}{N(p)}=
\gpinn (t)\bar{u}(p')i\gamma_{5}\tau_\alpha u(p)\\
\quad\\
\matel{N(p')}{\Gamma_{\sigma}}{N(p)}=
-\gsnn(t)\bar{u}(p') u(p)\\
\quad\\
\matel{N(p')}{\Gamma_{\omega}^{\mu}   }{N(p)}=
\bar{u}(p')\left[\gamma^\mu\gomegann(t)+\ds\frac{i}{2M}\fomegann(t)
\sigma^{\mu\nu}q_\nu\right]u(p)\\
\quad\\
\matel{N(p')}{\Gamma_{\rho}^{\alpha,\mu}   }{N(p)}=
\bar{u}(p')\tau_{a}\left[\gamma^\mu\grhonn(t)+\ds\frac{i}{2M}\frhonn(t)
\sigma^{\mu\nu}q_\nu\right] u(p)\\
\label{ff}
\ea
\eeq
where $\gpinn (t)$, $\gsnn(t)$, $\gomegann(t)$  $(\fomegann(t))$
and
$\grhonn(t)$  $(\frhonn(t))$ are form factors of
 $\pi-, \sigma-, \omega-, \rho-$
$NN$ vertices respectively; $t=q^2=q_{0}^{2}-\vec{q}^2=(p-p')^2$ is
the 4-momentum transfer and $M$ is the  nucleon mass.

In order to investigate the $T$ dependence we calculate three
line vertex correction, as is illustrated in Fig. 1 for
e.g. $\gpinn (t,T)$.
In accordance with Feynman rules this may be written as:
\bea
\ba{l}
\Gamma_A(t,T)=\Gamma_A(t)+i\ds\sum_B\Lambda_{AB}(t,T),\\
\quad \\
\Lambda_{AB}(t,T)=\ds\int\frac{d^4k}{(2\pi)^4}
\Gamma_B(k^2)S_F(a)\Gamma_A(t)S_F(b)\Gamma_B(k^2)\delb(k^2)
\label{gamma}
\ea
\eea
where $A, B$ - denote $\pi,\sigma,\omega,\rho$ mesons, e.g.
$\Gamma_{\pi}=\gpinn(t)$, $a=p'-k, b=p-k$. Here $S_F(a)$ and $\Delta^{B}(k^2)$
are the in - medium nucleon and meson propagators respectively.
In the framework of TFD the thermal propagator has a $2*2$
matrix  structure, but only the $11$-component refers to the
physical field. So, we may use the following representation:

\bea
\ba{l}
S_{F}^{11}(a)\equiv S_F(a)=(\hat a+M)[G_0(a)+G_T(a)]\,,\\
\quad \\
\Delta^{11}\equiv \Delta(k^{2})=D_0(k^2)+D_T(k^2)\,,\\
\quad \\
D_0(k^2)=\ds\frac{1}{k^2-m^2+i\varepsilon}\,,\quad
D_T(k^2)=-2\pi i n_B(k)\delta(k^2-m^2)\,,\\
G_0(a^2)=\ds\frac{1}{a^2-M^2+i\varepsilon}\,,\quad
G_T(a^2)=2\pi i \delta(a^2-M^2)N_F(a)\,,
\label{prop}
\ea
\eea
where $N_F(a)=\theta(a_0)n_F(a)+\theta(-a_0)\bar n_F(a)$,
$\theta(a_0)$ is the step function, $m$ is the mass
of corresponding meson and
\beq
n_F(a)=\frac{1}{e^{\beta(|a_0|-\mu)}+1}\,,\qquad
\bar n_F(a)=\frac{1}{e^{\beta(|a_0|+\mu)}+1}\,,\qquad
n_{B}(k)=\frac{1}{e^{\beta|k_0|}-1}
\label{nf}
\eeq
are the fermion, antifermion and boson distribution
functions respectively.
It is clear that when one substitutes Eq. (14)
into Eq. (13),  $\Lambda_{AB}(t,T)$ will
 be separated into two parts:
one refers to the naive zero temperature  contribution,
and the other one depends on $T$ and  $ \rho$ [4].
 We shall concentrate on the latter part, which may be rewritten
as follows:
\bea
\ba{l}
\Lambda_{AB}(t,T)=\ds\int\dsf{dk^4}{(2\pi)^4}\Gamma_B(k^2)
(\hat a+M)\Gamma_A(t)(\hat b+M)\Gamma_B(k^2)\times\\
\quad \\
\times[G_0(a)G_0(b)D_T^B(k^2)+
2G_0(a)G_T(b)D_0^B(k^2)]
\label{lab}
\ea
\eea
Note that we have neglected terms that are quadratic in the
temperature dependent distribution function. Now, using (16) and (14) in
(13) we get the following
expressions for the form factors:
\bea
\ba{l}
G_{ANN}(t,T)/G_{ANN}(t,T=0)=1+
i\ds\sum_B\int\dsf{dk^4}{(2\pi)^4}W_{AB}(t,k^2,T)
\quad \\
\times[G_0(a)G_0(b)D_T^B(k^2)+
2G_0(a)G_T(b)D_0^B(k^2)]
\label{final}
\ea
\eea
where the explicit formulas for $W_{AB}(t,k^2,T)$  may
 be found in the Appendix.
Here  we write expressions for
$W_{\omega\pi}(t,k^2,T)  $ and $W_{\rho\pi}(t,k^2,T)  $:
\bea
\ba{l}
W_{\omega\pi}(t,k^2,T) =\dsf{ 3G_{\pi NN}^2(k^2) } { 4 }
\left\{2[2M^2-(ab)]-
\dsf{3F_{\omega NN}(t)t} {2G_{\omega NN}(t)}\right\}\,;\quad\\
\quad\\
W_{\rho\pi}(t,k^2,T) =\dsf{-G_{\pi NN}^2(k^2)}{4}
\left\{2[2M^2-(ab)]-
\dsf{3F_{\rho NN}(t)t}{2G_{\rho NN}(t)}\right\}\,;\quad
\label{wpi}
\ea
\eea
where  $W_{\omega\pi}(t,k^2,T)$ and $W_{\rho \pi} (t, k^{2}, T)$
denote the contribution from the $\pi$-
exchange diagram to the $\gomegann(t,T)$ and $G_{\rho NN}(t, T)$
respectively.

\section{Results and discussions}

In order to start the  calculations, a set of free space meson - nucleon
form factors are chosen. We choose the  OBE monopole form factors
[8] (Bonn A):
 $G_{BNN}(t)=g_{BNN}(\Lambda_{B}^{2}-m_{B}^{2})/(\Lambda_{B}^{2}-t)$,
where , for example, $g_{\pi NN}^{2}/4\pi=14.09$ , and
 $\Lambda_{\pi NN}=1005MeV$.
Let us first discuss  the $T$ dependence of meson - nucleon coupling
constants - $g_{BNN}(T)\equiv G_{BNN}(t=m_{B}^{2},T)$.
The variation of the coupling constants $g_{\pi NN}(T)/g_{\pi NN}(T=0)$,
$g_{\sigma NN}(T)/g_{\sigma NN}(T=0)$,
$g_{\omega NN}(T)/g_{\omega NN}(T=0)$ and
$g_{\rho NN}(T)/g_{\rho NN}(T=0)$
with temperature are displayed in Fig. 2.
It is clear that they are nearly independent
of  the temperature below  $T_{c}^{B}$, then change rapidly for $T >
T_{c}^{B}$.
A similar behavior was also  predicted by Zhang et al. [4]
and by Dominguez et al.
[1] for pion - nucleon coupling constant
 $g_{\pi NN}(T)/g_{\pi NN}(T=0)$.
But here in Fig. 2,
an unexpected result is that, the $\omega -N$ coupling constant,
$g_{\omega NN}(T)$ increases
while the coupling constants of all other mesons decrease!
Let's consider a possible origin
of this controversy, by comparing the medium modification
of $\omega NN$ and $\rho NN$ coupling constants. In the present model
the modifications arise from
the triangle diagrams (Fig. 1).
Actual calculations show that, for any $G_{B NN}(t,T)$ the triangle
diagram with pion exchange gives a dominant contribution.
It is clear from Eq. (18) that at small $t$ the explicit
expressions for $g_{\omega NN}(T)$ and  $g_{\rho NN}(T)$
 formally coincide. However, $W_{\rho\pi}=-\frac{ W_{\omega\pi} }{3}$.
and hence the two have an opposite sign. The factor $(-1/3)$ arises
from the
isotopic spin structure. In fact, using  Eq. (13)
and Fig.1, $g_{\omega NN}(T)$ and  $g_{\rho NN}(T)$
may be written in a schematic way:
\beq
\ba{l}
g_{\omega NN}(T)\approx g_{\omega NN}(T=0)+
 g_{\pi NN}^{2} \ds\sum_{\beta}  \tau_{\beta}
                   g_{\omega NN}\tau_{\beta} +\dots\\
\quad\\
g_{\rho NN}(T)\tau_\alpha\approx g_{\rho NN}(T=0)\tau_\alpha+
 g_{\pi NN}^{2}\ds\sum_{\beta}  \tau_{\beta}\tau_\alpha
         g_{\rho NN}\tau_{\beta} +\dots
\ea
\eeq
where $\alpha$ and $\beta$ are isospin indices.
We omit the spin variables, since $\omega$ and $\rho$
have the same spin structure.
Now, using $\ds\sum_{\beta}\tau_{\beta}^2=3$ and
$\ds\sum_{\beta}\tau_{\beta}\tau_\alpha\tau_{\beta}=-\tau_\alpha$,
it is clear that the contribution of the leading
triangle diagram with pion exchange to   $g_{\omega NN}(T)$
and  $g_{\rho NN}(T)$ has different signs. In other words,
the term $\Lambda_{\rho\pi}$ in Eq. (13) is negative, while
$\Lambda_{\omega\pi}$   is positive.

In general, we conclude  from the results (Fig. 2)
that at a certain critical temperature $T_c^{B}$,
the couplings $g_{BNN}$ change drastically.
The thermal behavior of $g_{\pi NN} (T) = \gpinn(t,T)|_{t\to 0}$
has been investigated earlier [1] and we get here a similar behaviour of
$g_{\pi NN} (T)$
as a function of temperature. The temperature $T_c$, where $ g_{\pi NN} (T) $
changes dramatically,  was interpreted in [1] as a signal for the quark - gluon
deconfinement phase transition. Indeed, near $T_c$
the associated mean square radius
 $<r^{2}_{BNN}> = 6[\frac{ \partial}{\partial t} \ln G_{BNN}(t,T)]|_
{t \to 0}  $
is a monotonically increasing function of $T$ and in fact diverges
at the critical temperature. A similar behavior of
$<r^{2}_{BNN}>$ is also found in our calculations
and is illustrated in Fig. 3.
 As the critical temperature, $T_{c}^{B}$, is approached the
 strength of coupling of $\pi, \sigma$ and $\rho$ mesons
to nucleons is quenched, at the same time, the size of nucleons
as probed by the appropriate meson gets bigger. However,
 it is to be stressed that
$T_{c}^{B}$ is not the same for all mesons:
$T_{c}^{\pi} \approx 360 MeV$,
$T_{c}^{\sigma} \approx 95 MeV$,
$T_{c}^{\omega} \approx 175 MeV$ and
$T_{c}^{\rho} \approx 200 MeV$.
In OBE picture this means that , in the temperature region
e.g. $200 MeV< T <300 MeV $, the $\rho$  meson exchange is no longer
important while the pion exchange is still important.
On the other hand, the $\sigma$ and $\pi$ mesons are
mainly responsible for the attraction between nucleons. So,
the quenching of the $\sigma NN$ coupling constant
at $T>T_{c}^{\sigma}$ leads to the vanishing of the bound state, as it was
predicted earlier [4].

The density dependence of vertex form factors
has been studied in greater detail [2, 8]. It was found
that most of the vertex form factors
are quenched at high densities and it was anticipated that the
temperature dependence
is likely to yield results that are qualitatively similar to those of
density dependence.
   Now, let's consider their temperature dependence in some detail.
The form factors as a function of momentum transfer at several temperatures
are displayed in Fig. 4. It is seen that, the in-medium effects
lead to the suppression
of $\gpinn(t,T)$,  $\gsnn(t,T)$ and  $\grhonn(t,T)$.
The temperature dependence of $\gomegann (t,T)$,
is opposite to that of
$\grhonn$ due to it's isotopic structure as outlined above.

For practical calculations a parametrization
of these $G_{BNN}(t,T,\rho)$ form factors is needed.
At small momentum transfer   we can
 parametrize them by a monopole form :
\bea
\ba{l}
G_{BNN}(t,T,\rho)=g_{B}(T,\rho)(\Lambda_{B}^{2}(T,\rho)-m_{B}^{2})
/(\Lambda_{B}^{2}(T,\rho)+t)
\label{param}
\ea
\eea
 where in general the
 effective mass of a meson,
$m_{B}$, is also temperature dependent. Consideration of
this dependence is beyond the scope of the present paper. It will be
carried out
in a self consistent way, taking into account both the self energy
and the triangle diagrams, in a forthcoming paper.
But here, for simplicity, one may consider this as
 just a parametrization  and
choose the parameters $g_{B}(T,\rho)$
and $\Lambda_{B}^{2}(T,\rho)$.
Their  temperature dependence is still unknown.
Here  the T dependence  maybe represented, for $T<< T_{c}^{B}$, in a polynomial
form as:
\bea
\ba{l}
\dsf{g_{B}(T,\rho)}{g_{B}(T=0,\rho=0)}=1-\ds\alpha_B^g(\rho)(T/T_c^B)^2-
\ds\beta_B^g(\rho)(T/T_c^B)^4\\
\quad\\
\dsf{\Lambda_{B}(T,\rho)}{\Lambda_{B}(T=0,\rho=0)}=
1-\alpha_B^\lambda(\rho)(T/T_c^B)^2-
\beta_B^\lambda(\rho)(T/T_c^B)^4\\
\ea
\label{parg}
\eea
The calculated form factors
are fitted to this form and the paramters $\alpha$ and $\beta$ are
determined.
The results are presented in tables I, II, and III for
densities $\rho=0$,  $\rho=\rho_0$ and
$\rho=3\rho_0$, respectively (with $\rho_0=0.17fm^{-3}$ - the density
of normal nuclear matter). The results show that
the density dependence
of parameters, and hence, the form factors is rather weak.
What is the reason for such a result? The answer may be found by
considering expression in Eqs.
(14) - (17). It is clear that
the density, $\rho$, influences only through
the chemical potential $\mu$ defined from the particle and
antiparticle distribution functions as
\be
\rho=\ds\frac{4}{(2\pi)^3}\int d\vec k[n_F(k)-\bar n_F(k)],
\ee
where $\mu$ is involved in Eq. (15) in the exponent.
The argument of this exponent is $\beta(|a_0|\pm\mu)$.
Since $|a_0|\approx M=939 MeV$ is much larger
at moderate densities than $\mu$ ($\mu\approx 40 MeV $ for
$\rho=3\rho_0$, $T=150 MeV$),  the result is weakly sensitive
to $\mu$.

In summary, we have considered the temperature dependence
of meson - nucleon form factors and coupling constants in TFD formalism.
It is shown that at a critical temperature, where the coupling constant
changes drastically and the associated mean square radius diverges
is different for different mesons.
The temperature dependent vertex form factors are parametrized
in a simple monopole form and the T-dependence of
these parameters is clarified. These form factors
may be used in the calculation of the in - medium NN cross sections
[9]
and in investigations of the properties of hot dense matter.

In the present calculations,
the monopole form factors with the parameters (at T = 0)
given by the Bonn group, [7] are used. But the next question
which may arise is: are the results sensitive
to the shape of the input form factors? To answer this question
a different set of form factors
 obtained by Meissner et. al., [10] in the framework
of  a topological soliton model have been used.
The thermal behaviour of such form factors  is shown in
Fig. 5. Now, comparing Fig. 4 and Fig. 5 it is clear
that only the $\sigma NN$ form factor is affected the most.
The reason is that even in free space and at
$T=0$ the $\sigma NN$ form factor in the topological soliton model [10]
is much  harder ($\Lambda_{\sigma}\approx 2700 MeV$)
 than those of the Bonn potential ($\Lambda_{\sigma}\approx 2000 MeV$).
Nevertheless our main conclusions about thermal behavior
of different meson nucleon vertices remain valid.

The variation of the meson-nucleon coupling constant with temperature
(Fig. 1) clearly indicates that the attractive NN force due to
$\sigma$-exchange decreases quite rapidly while the repulsive N-N
force due to $\omega$-exchanges increases. This would lead to the
fact that the nuclear matter is quite likely un-bound at high
temperature. In fact it will look more like a hard sphere gas since the
repulsive interaction at short distances will dominate. The attractive
interactions in such a case play only a small perturbative role. The
hot and dense nuclear matter could be approximated as a free gas with
an excluded volume around each nucleon. Perhaps the extensive studies
[11]
along this line can be justified on the basis of the results obtained
in this paper. The nuclear matter would be hard to compress.

It would be quite interesting to know, if the critical temperature,
$T_{c}^{B}$, maybe uniquely considered as
a point of transition from the hadron state to a quark gluon state,
as it was suggested [1].
The usual way of estimating $T_c$  is based on QCD lattice
calculations
[12] or temperature
 dependent quark - gluon potentials [13].
In such calculations it is usual to start from high temperatures
and then decreases the  temperature, looking for a
phase transition from the quark gluon state to a hadron state i.e.
hadronization of the quarks.
Clearly this method gives a unique $T_c$, since at any stage there is
either a hadron or a quark gluon plasma phase. Contrary to such studies
the present calculations explore the transition froma hadron to a
quark-gluon phase and find that the critical temperature for
hadron $\rightarrow$ quark-gluon
transition depends on the kind of hadrons
 under consideration. A similar result, where  hadrons and
quark - gluon plasma coexist have been obtained in lattice calculations
[12].
In conclusion it is important to emphasize that the dynamics of
hadrons and of quark-gluon plama at finite temperature is poorly
understood. The question of critical phenomenon, in particular
critical temperature, in the hadronic systems is not well-known.
Studies in nuclear matter are needed urgently
to clarify the critical transition from a hadronic system to the quark
gluon phase which will depend on both density and temperature. The
experiments with relativistic heavy ion colliders would need such an
understanding to clarify the production of quark-gluon plasma from
hadrons and the eventual hadronisation of the quark-gluon system. It
is possible that there will be clear and unambiguous signals for the
formation and subsequent hadronisation of the quark-gluon plasma [14].

\section*{Acknowledgements.}

We are indebted to Prof. M.M. Musakhanov, Prof. A. Nakamura and
Prof. W.M. Alberico
for their useful discussions. We thank Prof. C.P. Singh for a critical
reading of the manuscript. U. Yakhshiev acknowledges partial support
by
% the grant INTAS-93-0239ext and by
 the grant 11/97 of the State Committee
for Science and Technology of Uzbekistan. The research of F.C. Khanna
is supported in part by the National Sciences and Engineering
research council of Canada.

\newpage
\section*{Appendix}

Here  explicit expressions for $W_{AB}(t,k^2,T)$ introduced
in eq. (7) are given in detail.
\bea
\ba{l}
W_{\pi\pi}=G_{\pi}(k^2)[M^2-(ab)]\,; \quad
W_{\pi\sigma}=W_{\pi\pi}|_{G_{\pi}\rightarrow G_{\sigma}}\,;\\
W_{\pi\omega}=A_{\pi}^\omega[M^2-(ab)]+3F_1^\omega(k^2)F_2^\omega(k^2)
z\,; \quad
W_{\pi\rho}=-W_{\pi\omega}\left|_{\footnotesize\ba{l}
F_i^\omega\rightarrow F_i^\rho\\
A_\pi^\omega\rightarrow A_\pi^\rho
\ea}\right.\,;\\
A_{\pi}^\omega=[F_1^\omega(k^2)]^2[4-\ds\frac{k^2}{m_\omega^2}]+
\frac{3k^2[F_2^\omega(k^2)]^2}{4M^2}\,;\\
%%\quad \\
W_{\sigma\pi}=-3g_\pi^2(k^2)[M^2+(ab)]\,,\quad
W_{\sigma\sigma}=-\ds\frac{1}{3}W_{\sigma\pi}|_{G_\pi\rightarrow
G_\sigma}\,,\\
%%\quad\\
W_{\sigma\omega}=[M^2+(ab)]A_\sigma^\omega\,;\quad
W_{\sigma\rho}=3W_{\sigma\omega}|_{
A_\sigma^\omega\rightarrow A_\sigma^\rho}\,;\\
%%\quad \\
A_{\sigma}^\omega=-[F_1^\omega(k^2)]^2[4-\ds\frac{k^2}{m_\omega^2}]+
\frac{3k^2[F_2^\omega(k^2)]^2}{4M^2}\,;\\
%%\quad \\
W_{\omega\pi} =\dsf{3g_{\pi}^2(k^2)}{4}\left\{2[2M^2-(ab)]-
\dsf{3F_2^\omega(t)t}{2F_1^\omega(t)}\right\}\,;\quad
W_{\omega\sigma}=\dsf{1}{3}W_{\omega\pi}|{G_{\pi}\rightarrow G_\sigma}\,;\\
%%\quad \\
W_{\omega\omega}=\dsf{1}{4}\left\{2\Gamma_1^\omega[2M^2-(ab)]+
\dsf{\Gamma_2^\omega}{M^2}[k^2[M^2-(ab)]+2(ak)(bk)]+\right.\\
%%\quad \\
+3z \Gamma_3^\omega-\dsf{F_2^\omega(t)}{2M^2F_1^\omega(t)}
[3\Gamma_1^\omega t+\Gamma_2^\omega(k^2t-z^2)]+\\
%%\quad \\
+\left.\dsf{\Gamma_3^\omega}{M^2}[z[(ab)+3M^2]-
2(ak)(bq)-2(aq)(bk)]\right\}\,;\\
\quad \\
\Gamma_1^\omega=[F_1^\omega(k^2)]^2[2-\ds\frac{k^2}{m_\omega^2}]+
\frac{k^2[F_2^\omega(k^2)]^2}{4M^2},\,\quad
\quad\\
\Gamma_2^\omega=\dsf{2M^2}{m_\omega^2}[F_1^\omega(k^2)]^2
-[F_2^\omega]^2\,;
\quad\\
\Gamma_3^\omega=F_1^\omega(k^2) F_2^\omega(k^2)\,,\quad
\quad W_{\omega\rho}=3 W_{\omega\omega}|_{\Gamma_i^\omega\rightarrow
\Gamma_i^\rho}\,,\quad
W_{\rho\pi}=-\dsf{1}{3}W_{\omega\pi}|_{\omega\rightarrow\pi}\,;\\
%\nonumber
%\ea\eea
%\bea\ba{l}
W_{\rho\sigma}=-W_{\rho\pi}|_{G_\pi\rightarrow G_\sigma}\,;\quad
W_{\rho\omega}=W_{\omega\omega}\left|_{\footnotesize\ba{l}
\Gamma_i^\omega\rightarrow\Gamma_i^\omega\\
F_i^\omega\rightarrow F_i^\rho
\ea}\right.\,;\quad W_{\rho\rho}=-W_{\omega\omega}|_{\omega\rightarrow\rho}\,;
\nonumber
\ea\eea
where $G_{\pi}\equiv G_{\pi NN}(k^2)$,
$G_{\sigma}\equiv G_{\sigma NN}(k^2)$,
$F_1^\omega\equiv G_{\omega NN}$, $F_2^\omega\equiv F_{\omega NN}$
, $z=(kq)$ and $q^2=t$.

\newpage

\centerline {\bf FIGURE CAPTIONS}

\begin{description}
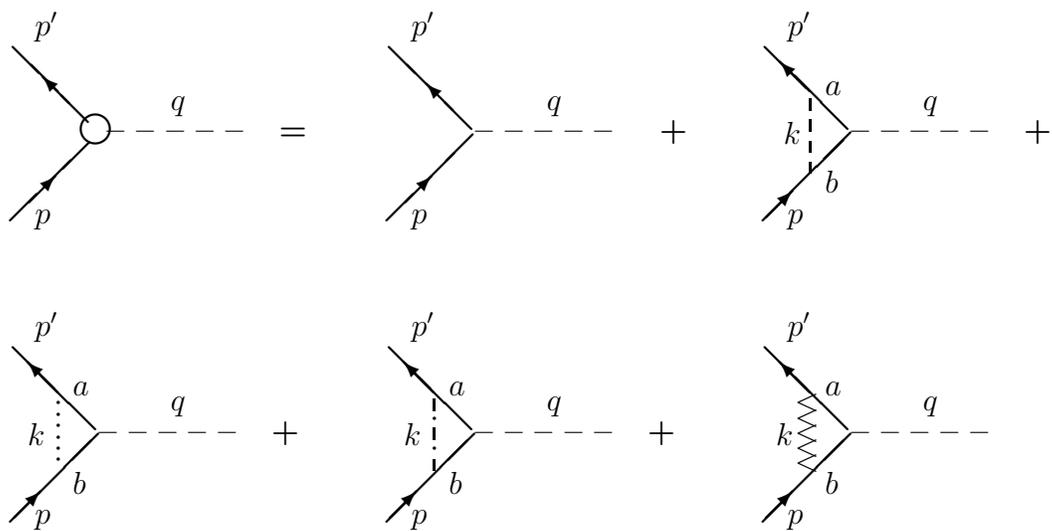
\item [Fig. 1.]
Feynman diagrams for pion - nucleon vertex, The solid line is for nucleon.
Dashed, dotted, dot - dashed and wavy lines for
$\pi$, $\sigma$, $\omega$ and  $\rho$ mesons.
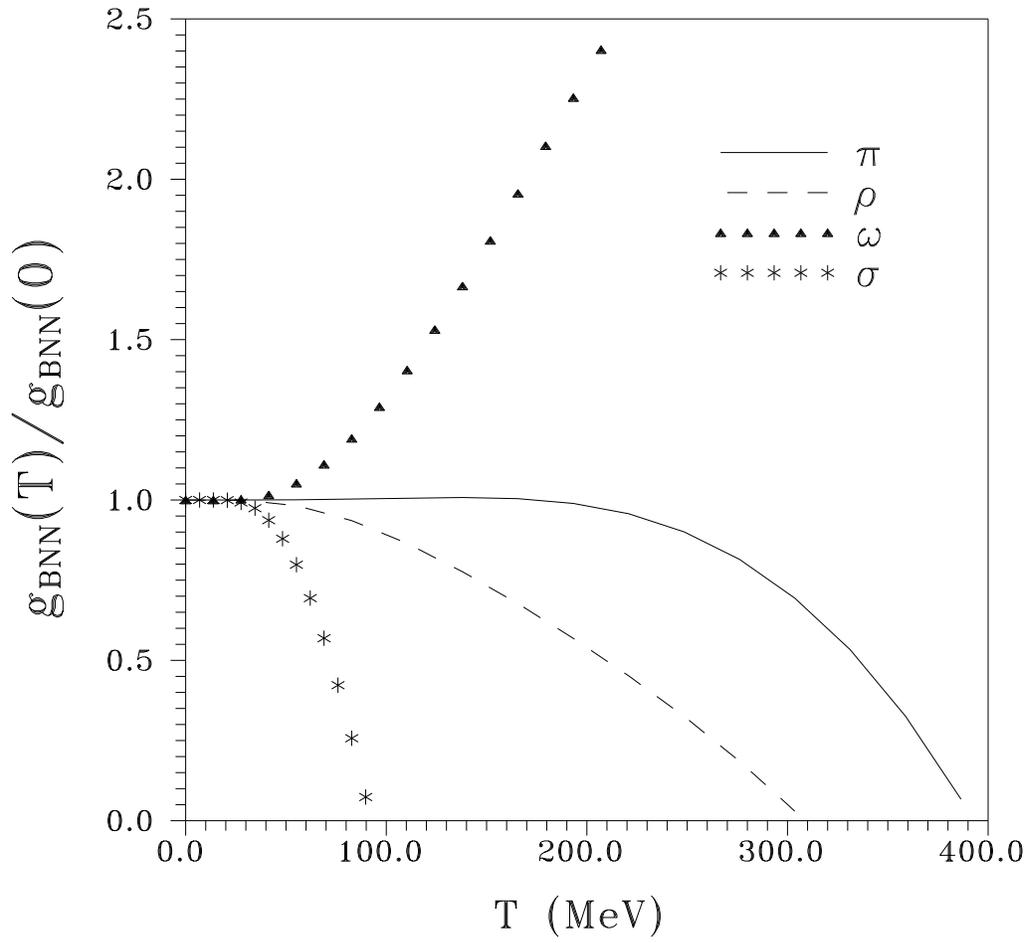
\item [Fig. 2.]
The ratio of meson-nucleon coupling constants at finite temperature T
and at T = 0 as a function of temperature, T.
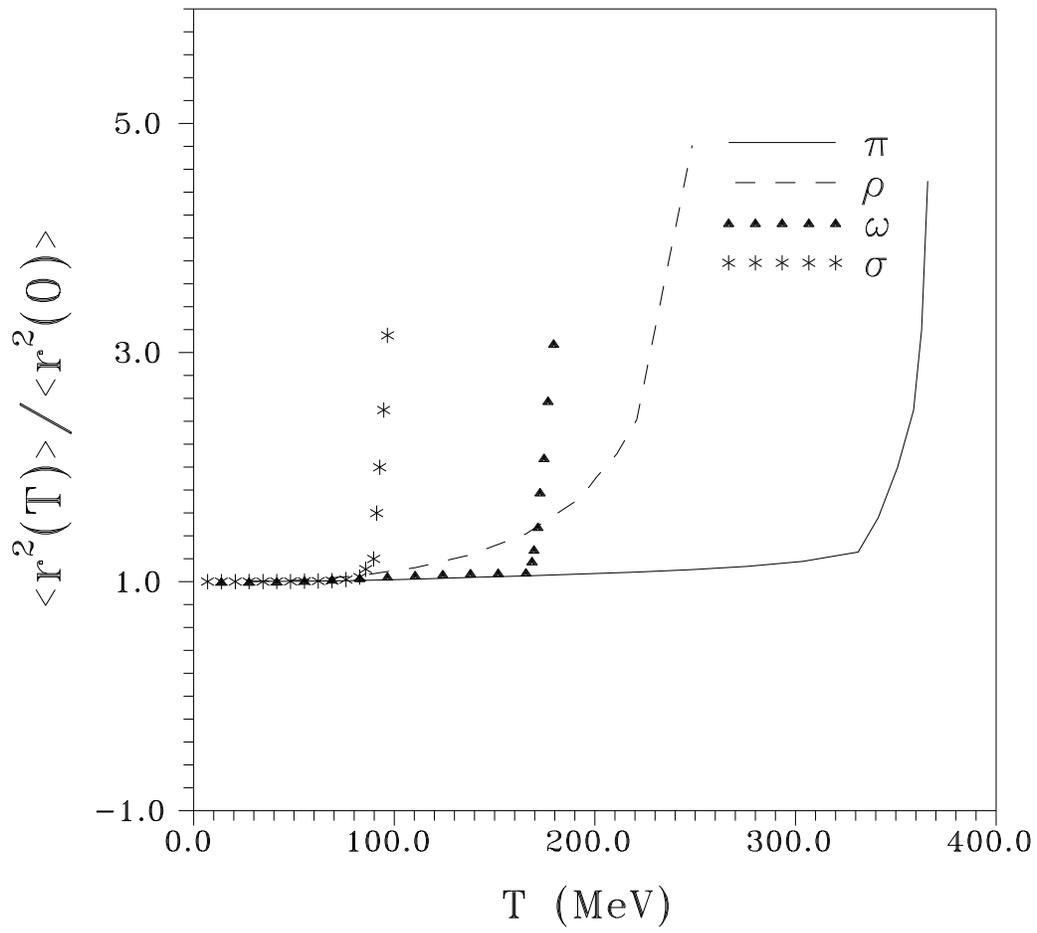
\item [Fig. 3.]
The ratio of mean square radius at finite temperature T and at T = 0
as a function of temperature, T.
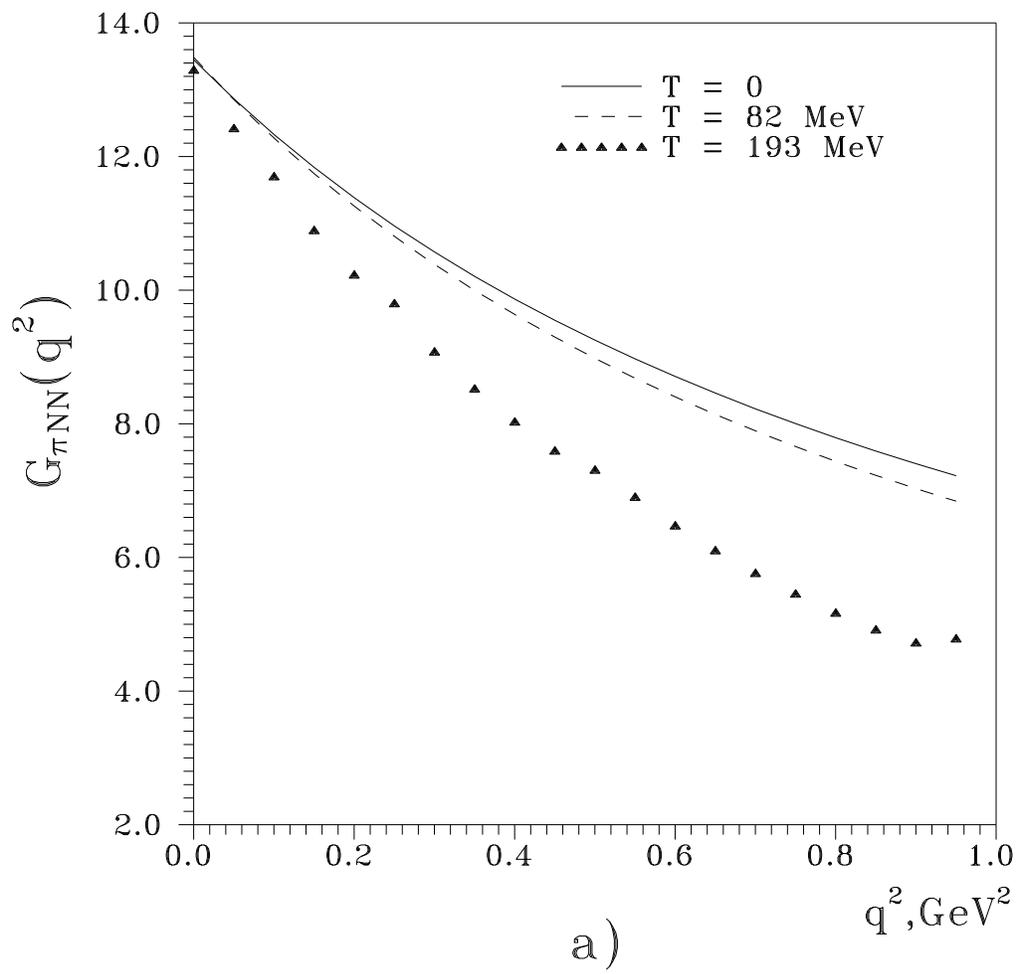
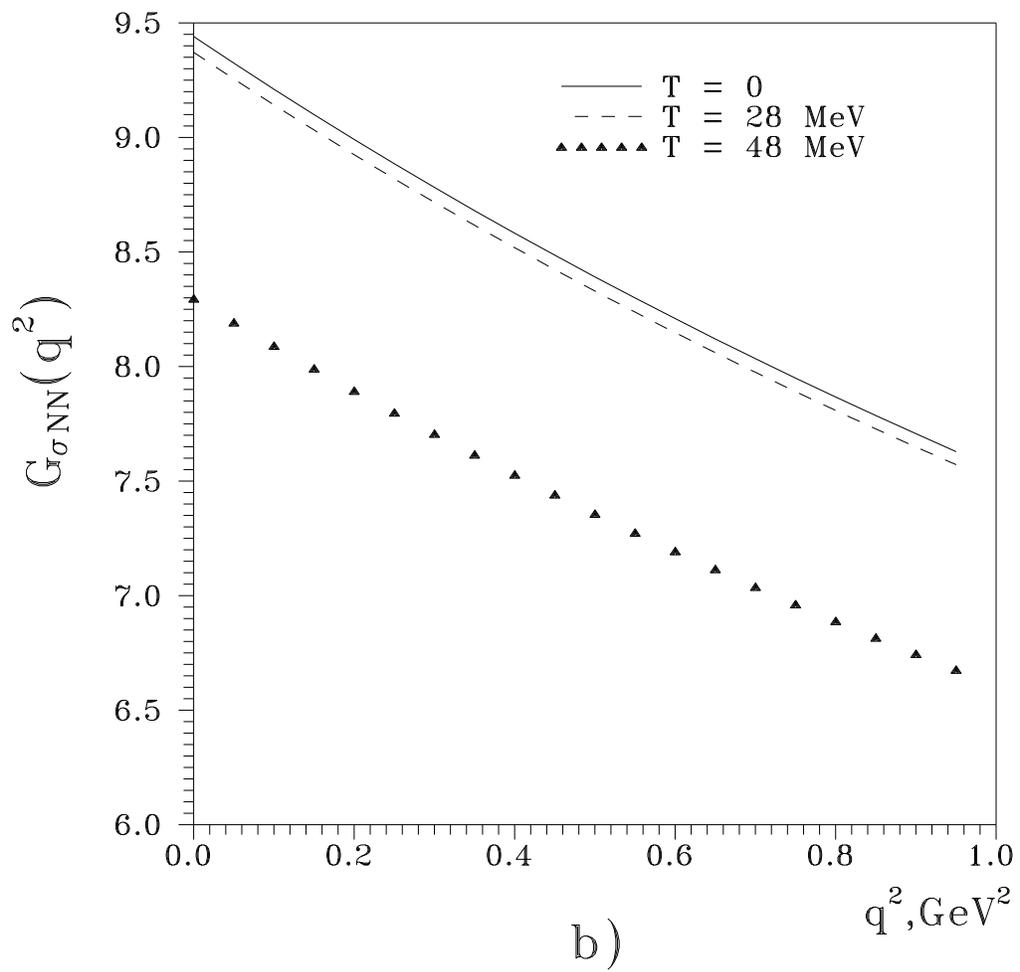
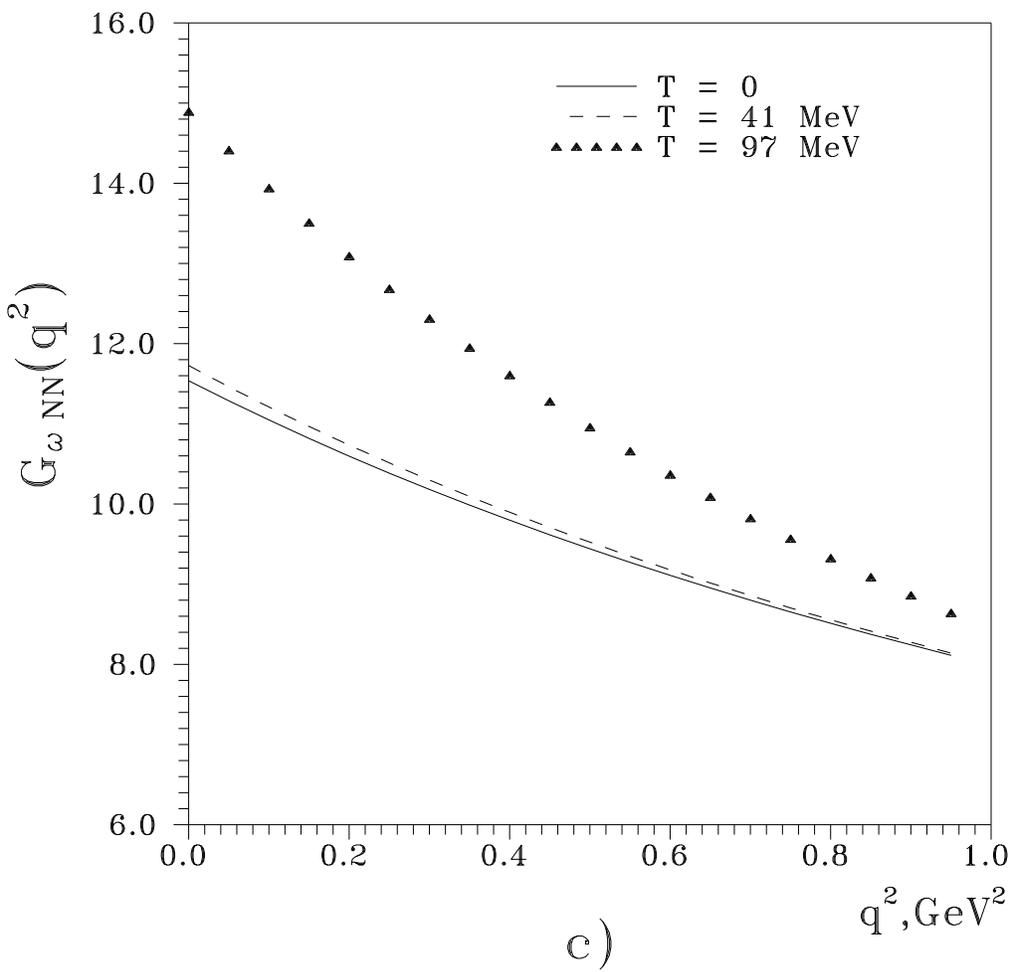
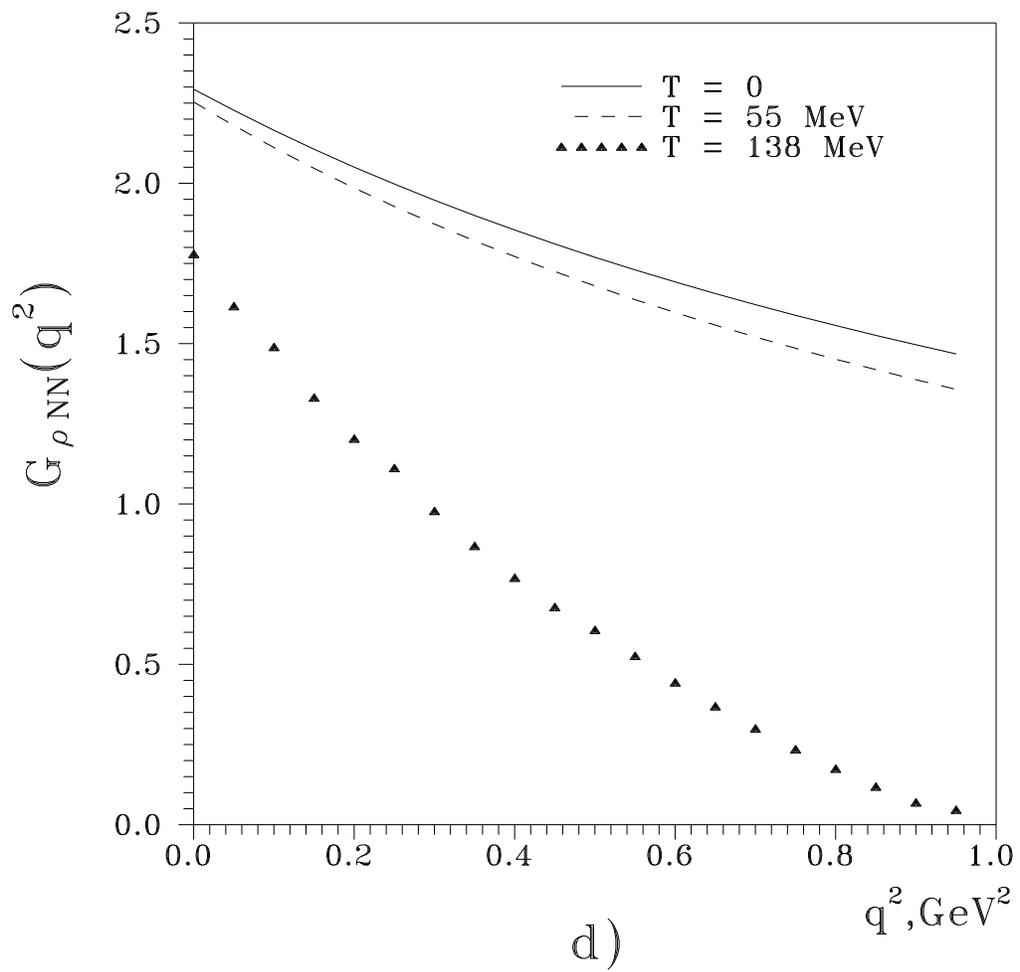
\item [Fig. 4.]
Meson nucleon form factors at several temperatures as a function of
$q^{2}$ using parametrisation of BONN group [7]. a) $\pi NN$ vertex;
b) $\sigma NN$ vertex; c) $\omega NN$ vertex and d) $\rho NN$ vertex.
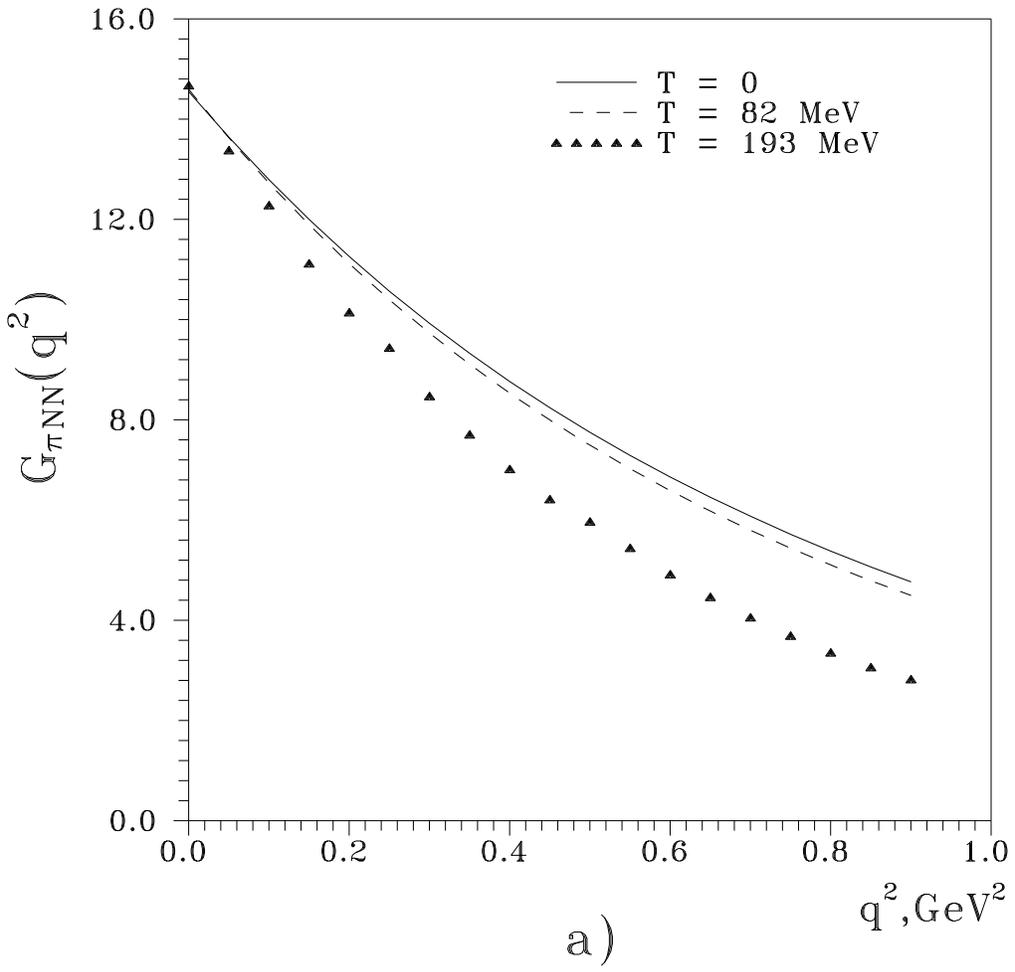
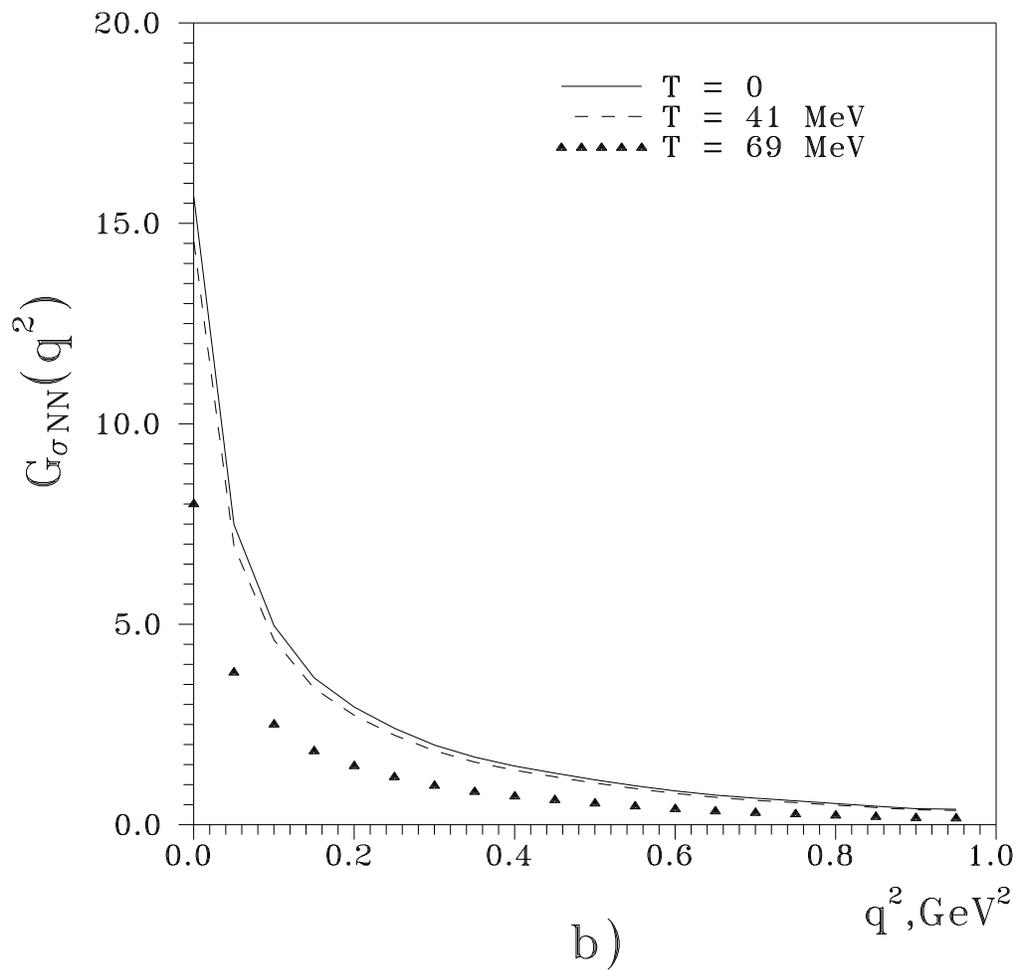
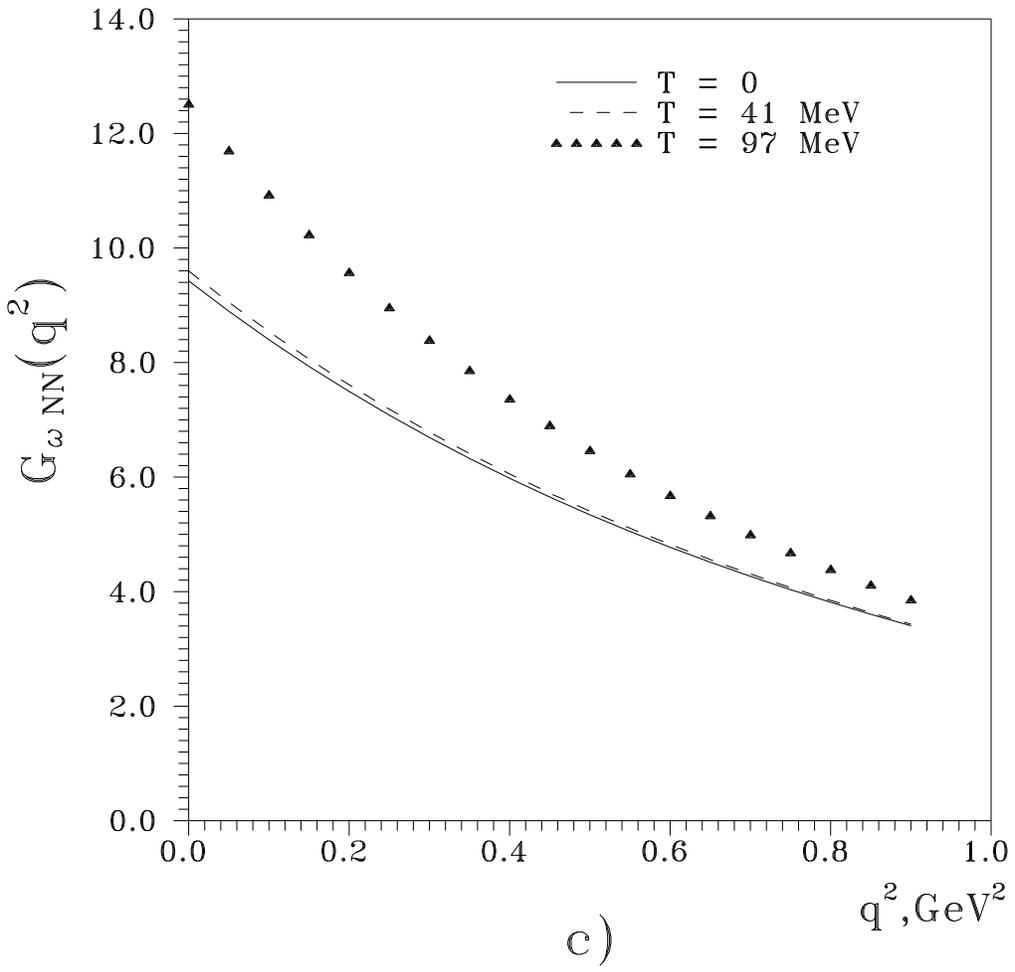
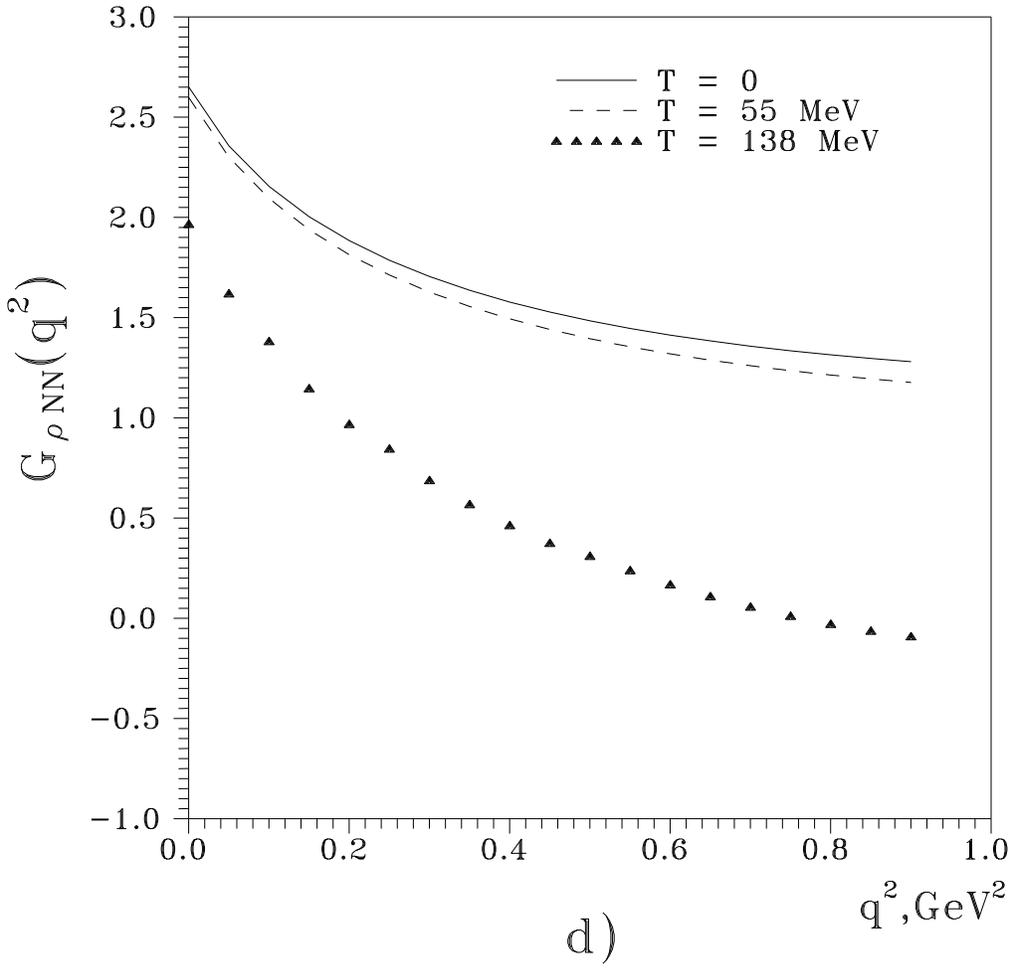
\item [Fig. 5.]
Meson-nucleon form factors at several temperatures as a function of
$q^{2}$ using parametrisation of Meissnar et al [10]. a) $\pi NN$
vertex; b) $\sigma NN$ vertex; c) $\omega NN$ vertex and d) $\rho NN$
vertex.
\end{description}

\newpage

\begin{table}
\caption{Parameters  of vertex form factors in Eq.s (20)-(21) at zero
density $\rho=0.$}
\vskip 0.2cm
\begin{tabular}{cccccc}
Meson&$\alpha_g$&$\beta_g$&$\alpha_\lambda$&$\beta_\lambda$&$T_c$\\
\hline
$\pi$   &-0.201& 0.884& 0.6237&-0.08419&$\approx 360$\\
$\sigma$& 0.3252& 0.8381&-0.3245& 0.9184&$\approx 95$\\
$\rho$  & 0.4384& 0.03194& 1.198&-0.5245&$\approx 200$\\
$\omega$&-0.8951&-0.2235& 0.7322&-0.5258&$\approx 175$\\
\end{tabular}
\end{table}

\begin{table}
\caption{The same as in Table I , but for $\rho=\rho_{0}$}
\vskip 0.2cm
\begin{tabular}{cccccc}
Meson&$\alpha_g$&$\beta_g$&$\alpha_\lambda$&$\beta_\lambda$&$T_c$\\
\hline
$\pi$   &-0.1966& 0.8899& 0.6181&-0.0750 &$\approx 360$\\
$\sigma$& 0.3254& 0.8378&-0.3301& 0.9329&$\approx 95$\\
$\rho$  & 0.4379& 0.03347& 1.2734&-0.6071&$\approx 200$\\
$\omega$&-0.8966&-0.2168& 0.7332&-0.5366&$\approx 175$\\
\end{tabular}
\end{table}
\begin{table}
\caption{The same as in Table I , but for $\rho=3\rho_{0}$}
\vskip 0.2cm
\begin{tabular}{cccccc}
Meson&$\alpha_g$&$\beta_g$&$\alpha_\lambda$&$\beta_\lambda$&$T_c$\\
\hline
$\pi$   &-0.1868& 0.9012& 0.6056 &-0.05452 &$\approx 360$\\
$\sigma$& 0.3258& 0.8366&-0.3429& 0.9674&$\approx 95$\\
$\rho$  & 0.4379& 0.03347& 1.1934&-0.5047&$\approx 200$\\
$\omega$&-0.8994&-0.2026& 0.7350&-0.5609&$\approx 175$\\
\end{tabular}
\end{table}

\newpage
\pagestyle{empty}
%%%%%%%%%%%%%%%%%%PICTURE%%%%%%%%%%%%%%%%%%%%%%%%%%%%%%%%%%%%%%%%
\begin{center}
\setlength{\unitlength}{0.1mm}
\begin{picture}(1500,800)
\thicklines
\normalsize
%%%%1
\put(250,600){$q$}
\put(100,625){\line(-1,1){60}}
\put(140,585){\vector(-1,1){60}}
\put(82,498){\line(1,1){60}}
\put(37,453){\vector(1,1){60}}
\put(150,575){\circle{40}}
\put(70,700){$p^{\prime}$}
\put(70,450){$p$}
%%%%2
\put(750,600){$q$}
\put(165,560){\bf -- -- -- -- --\quad  =}
\put(600,625){\line(-1,1){60}}
\put(650,575){\vector(-1,1){60}}
\put(592,508){\line(1,1){60}}
\put(537,453){\vector(1,1){60}}
\put(655,560){\bf -- -- -- -- -- \quad +}
\put(570,700){$p^{\prime}$}
\put(570,450){$p$}
%%%%3
\put(1250,600){$q$}
\put(1065,555){$k$}
\put(1100,625){\line(-1,1){60}}
\put(1150,575){\vector(-1,1){90}}
\put(1074,490){\line(1,1){80}}
\put(1037,453){\vector(1,1){40}}
\put(1155,560){\bf -- -- -- -- --\quad +}
\put(1070,700){$p^{\prime}$}
\put(1070,450){$p$}
\put(1120,620){$a$}
\put(1120,490){$b$}
\put(1100,600){\line(0,1){15}}
\put(1100,572){\line(0,1){15}}
\put(1100,544){\line(0,1){15}}
\put(1100,516){\line(0,1){15}}
%%%%4
\put(60,155){$k$}
\put(120,220){$a$}
\put(120,90){$b$}
\put(250,200){$q$}
\put(100,225){\line(-1,1){60}}
\put(150,175){\vector(-1,1){90}}
\put(74,90){\line(1,1){80}}
\put(37,53){\vector(1,1){40}}
\put(155,160){\bf -- -- -- -- --\quad +}
\put(70,300){$p^{\prime}$}
\put(70,50){$p$}
\put(95,200){$\cdot$}
\put(95,185){$\cdot$}
\put(95,170){$\cdot$}
\put(95,155){$\cdot$}
\put(95,140){$\cdot$}
\put(95,125){$\cdot$}
%%%%5
\put(560,155){$k$}
\put(620,220){$a$}
\put(620,90){$b$}
\put(750,200){$q$}
\put(600,225){\line(-1,1){60}}
\put(650,175){\vector(-1,1){90}}
\put(574,90){\line(1,1){80}}
\put(537,53){\vector(1,1){40}}
\put(655,160){\bf -- -- -- -- --\quad +}
\put(570,300){$p^{\prime}$}
\put(570,50){$p$}
\put(600,200){\line(0,1){15}}
\put(595,180){$\cdot$}
\put(600,162){\line(0,1){15}}
\put(595,140){$\cdot$}
\put(600,120){\line(0,1){15}}
%%%%6
\put(1055,155){$k$}
\put(1120,220){$a$}
\put(1120,90){$b$}
\put(1250,200){$q$}
\put(1100,225){\line(-1,1){60}}
\put(1150,175){\vector(-1,1){90}}
\put(1074,90){\line(1,1){80}}
\put(1037,53){\vector(1,1){40}}
\put(1155,160){\bf -- -- -- -- -- }
\put(1070,300){$p^{\prime}$}
\put(1070,50){$p$}
\put(1080,200){$<$}
\put(1080,180){$<$}
\put(1080,160){$<$}
\put(1080,140){$<$}
\put(1080,120){$<$}
\end{picture}
\end{center}
%%%%%%%%%%%%%%%%%%%%%%%%%%%%%%%%%%%%%%%%%%%%%%%%%%%%%%%%%%%%%%%%%

\begin{center}
Fig. 1
\end{center}

\begin{thebibliography}{99}
\bibitem{} C.A. Dominguez, C. van Gend and M. Loewe Phys. Lett.
               B429, 64, (1998).
\bibitem{} A.M. Rakhimov , F.C. Khanna, U.T. Yakhshiev and
                  M.M. Musakhanov Nucl. Phys. A643, 383, (1998).
\bibitem{}
  Y. Takahashi and H. Umezawa, Collective Phenomenon 2, 55, (1975) -
  reprinted in Int. J. Mod. Physics B10, 1755, (1996); H.  Umezawa et.  al.,
  Thermo Field Dynamics and condensed states  (North-
     Holland,  Amsterdam,  1982);  H.  Umezawa, Advanced Field
     Theory (American Institute of Physics, New York, 1993);
 P.A.  Henning, Phys. Rep. 253, 235, (1995).
\bibitem{} Yi - Jun Zhang, Song Gao   and Ru - Keng Su
               Phys. Rev. C56, 3336, (1997) and references there in.
\bibitem{}
 L. Alvarez -
      Ruso,  P.F.  de Cordoba and E. Oset, Nucl. Phys. A606,
      407, (1996); Y. Kim, H.K. Lee nucl-th/9609020.
\bibitem{} J.W. Harris, Proc. Lake Louise Winter Institute, Ed. A.
Astbury et. al., (World Scientific, 1999). This review contains
numerous references to both experimental and theoretical articles.
\bibitem{} R. Machleidt  Adv. Nucl. Phys. 19, 189, (1989).
\bibitem{}
   Ulf-G. Meissner, Nucl. Phys. A503, 801, (1989).
\bibitem{} G.Q. Li and R. Machelidt Phys. Rev. C 48, 1702, (1993).
\bibitem{} U. Meissner, A. Rakhimov, U. Yakhshiev
 nucl-th/9901067.
\bibitem{} R. Hagedorn and J. Rafelski, Phys. Lett. B97, 136, (1980); R.
Hagedorn, Z. Phys. C17, 265, (1985); C.P. Singh, B.K. Patra and K.K.
Singh, Phys. Lett. B387, 680, (1996); J. Sollfrank, J. Phys. G23,
1903, (1997).
\bibitem{} H. Satz talk at Int. Conf. on Physics and Astrophysics of
Quark-Gluon plasma, Ed. B.C. Sinha et. al., (Naros publications, 1998).
\bibitem{alberico} W.M. Alberico, M. Nardi and S. Quatrocolo
                    Nucl. Phys. A589 , 620, 1995.

\bibitem{}C.P. Singh, Phys. Rep. 236, 147 (1993); S.A. Bass, M.
Gyulassy, H. Stoker and W. Greiner, J. Phys. G25, R1 (1999) and the
references therein.
\end{thebibliography}
\end{document}